\documentclass[letterpaper,twocolumn,superscriptaddress,floatfix]{revtex4}
\usepackage[latin1]{inputenc}
\usepackage{bm}
\usepackage{multirow,amssymb,amsbsy,amsmath}
\usepackage{graphicx}
\usepackage{verbatim}
\usepackage{braket}
\makeatletter
\usepackage{pifont}
\makeatother

\usepackage{graphicx}
\usepackage{dcolumn}
\usepackage{bm}
\usepackage{ifthen}
\usepackage{booktabs}
\usepackage{nicefrac}
\usepackage{float}
\usepackage{color,xcolor}

\usepackage{caption}
\begin{document}

\title{Temperature-dependent energy-level shifts of spin defects in hexagonal boron nitride}
\author{Wei Liu}
\affiliation{CAS Key Laboratory of Quantum Information, University of Science and Technology of China, Hefei, P.R.China}
\affiliation{CAS Center For Excellence in Quantum Information and Quantum Physics, University of Science and Technology of China, Hefei, P.R.China}
\author{Zhi-Peng Li}
\affiliation{CAS Key Laboratory of Quantum Information, University of Science and Technology of China, Hefei, P.R.China}
\affiliation{CAS Center For Excellence in Quantum Information and Quantum Physics, University of Science and Technology of China, Hefei, P.R.China}
\author{Yuan-Ze Yang}
\affiliation{CAS Key Laboratory of Quantum Information, University of Science and Technology of China, Hefei, P.R.China}
\affiliation{CAS Center For Excellence in Quantum Information and Quantum Physics, University of Science and Technology of China, Hefei, P.R.China}
\author{Shang Yu}
\affiliation{CAS Key Laboratory of Quantum Information, University of Science and Technology of China, Hefei, P.R.China}
\affiliation{CAS Center For Excellence in Quantum Information and Quantum Physics, University of Science and Technology of China, Hefei, P.R.China}
\author{Yu Meng}
\affiliation{CAS Key Laboratory of Quantum Information, University of Science and Technology of China, Hefei, P.R.China}
\affiliation{CAS Center For Excellence in Quantum Information and Quantum Physics, University of Science and Technology of China, Hefei, P.R.China}
\author{Zhao-An Wang}
\affiliation{CAS Key Laboratory of Quantum Information, University of Science and Technology of China, Hefei, P.R.China}
\affiliation{CAS Center For Excellence in Quantum Information and Quantum Physics, University of Science and Technology of China, Hefei, P.R.China}
\author{Ze-Cheng Li}
\affiliation{CAS Key Laboratory of Quantum Information, University of Science and Technology of China, Hefei, P.R.China}
\affiliation{CAS Center For Excellence in Quantum Information and Quantum Physics, University of Science and Technology of China, Hefei, P.R.China}
\author{Nai-Jie Guo}
\affiliation{CAS Key Laboratory of Quantum Information, University of Science and Technology of China, Hefei, P.R.China}
\affiliation{CAS Center For Excellence in Quantum Information and Quantum Physics, University of Science and Technology of China, Hefei, P.R.China}
\author{Fei-Fei Yan}
\affiliation{CAS Key Laboratory of Quantum Information, University of Science and Technology of China, Hefei, P.R.China}
\affiliation{CAS Center For Excellence in Quantum Information and Quantum Physics, University of Science and Technology of China, Hefei, P.R.China}
\author{Qiang Li}
\affiliation{CAS Key Laboratory of Quantum Information, University of Science and Technology of China, Hefei, P.R.China}
\affiliation{CAS Center For Excellence in Quantum Information and Quantum Physics, University of Science and Technology of China, Hefei, P.R.China}
\author{Jun-Feng Wang}
\affiliation{CAS Key Laboratory of Quantum Information, University of Science and Technology of China, Hefei, P.R.China}
\affiliation{CAS Center For Excellence in Quantum Information and Quantum Physics, University of Science and Technology of China, Hefei, P.R.China}
\author{Jin-Shi Xu}
\affiliation{CAS Key Laboratory of Quantum Information, University of Science and Technology of China, Hefei, P.R.China}
\affiliation{CAS Center For Excellence in Quantum Information and Quantum Physics, University of Science and Technology of China, Hefei, P.R.China}
\author{Yi-Tao Wang}
\email{yitao@ustc.edu.cn}
\affiliation{CAS Key Laboratory of Quantum Information, University of Science and Technology of China, Hefei, P.R.China}
\affiliation{CAS Center For Excellence in Quantum Information and Quantum Physics, University of Science and Technology of China, Hefei, P.R.China}
\author{Jian-Shun Tang}
\email{tjs@ustc.edu.cn}
\affiliation{CAS Key Laboratory of Quantum Information, University of Science and Technology of China, Hefei, P.R.China}
\affiliation{CAS Center For Excellence in Quantum Information and Quantum Physics, University of Science and Technology of China, Hefei, P.R.China}
\author{Chuan-Feng Li}
\email{cfli@ustc.edu.cn}
\affiliation{CAS Key Laboratory of Quantum Information, University of Science and Technology of China, Hefei, P.R.China}
\affiliation{CAS Center For Excellence in Quantum Information and Quantum Physics, University of Science and Technology of China, Hefei, P.R.China}
\author{Guang-Can Guo}
\affiliation{CAS Key Laboratory of Quantum Information, University of Science and Technology of China, Hefei, P.R.China}
\affiliation{CAS Center For Excellence in Quantum Information and Quantum Physics, University of Science and Technology of China, Hefei, P.R.China}

\date{\today }
\begin{abstract}

\textbf{Abstract:} Two-dimensional hexagonal boron nitride (hBN) has attracted much attention as a platform for realizing integrated nanophotonics, and collective effort has been focused on spin defect centers. Here, the temperature dependence of the optically detected magnetic resonance (ODMR) spectrum of negatively charged boron vacancy (V$_\text{B}^-$) ensembles in the range of 5-600 K is investigated. The microwave transition energy is found to decrease monotonically with increasing temperature and can be described by the Varshni empirical equation very well. Considering the proportional relation between energy-level shifts and the reciprocal lattice volume ($V^{-1}$), thermal expansion might be the dominant cause for energy-level shifts. We also demonstrate that the V$_\text{B}^-$ defects are stable at up to 600 K. Moreover, we find that there are evident differences among different hBN nanopowders, which might originate from the local strain and distance of defects from the flake edges. Our results may provide insight into the spin properties of V$_\text{B}^-$ and for the realization of miniaturized, integrated thermal sensors.

\end{abstract}


\maketitle 
\bibliographystyle{prsty}

Optically polarizing spin states of color centers in solids play a key role in developing the route toward the realization of quantum information \cite{Lambropoulos2007} and sensing \cite{Degen2017} applications. Significant effort has been  devoted to discovering and assessing quantum properties of spin defects, including the nitrogen-vacancy (NV) center \cite{Jelezko2006, Acosta2010, ChenXD2011, Toyli2012, Doherty2014, Ivady2014} and the silicon-vacancy center \cite{Sipahigil2016, Rose2018} in diamond, the divacancy center \cite{Koehi2011, ZhouY2017, YanFF2018, YanFF2020, QLi2020} and the nitrogen-vacancy center \cite{Hijikata2019, JFWang2020} in silicon carbide, and the rare-earth impurities in complex oxides \cite{Siyushev2014}. Although these defects have many exceptional features, such as long coherence times and single spin manipulation, there is still huge interest in identifying similar defects in other materials, as they may offer an expanded range of functionality.

Hexagonal boron nitride (hBN) is a wide-band-gap ($\sim$6 eV) two-dimensional (2D) semiconductor with some excellent properties. It is very useful and promising in many applications due to its unique optoelectronic and nanophotonic effects \cite{Caldwell2019}. For example, hBN has been suggested as a core material for thermal management in electronics due to its exceptionally high thermal conductivity \cite{Sichel1976, Jo2013, ZhengJC2016}; some specific heterostructures composed of hBN and other 2D materials are useful systems to study moir\'e physics \cite{ChenG2019N, ChenG2019NP, TangY2020}; and the strongly confined hyperbolic phonon-polariton modes in hBN have led to a series of applications in nanophotonics \cite{DaiS2015, LiP2015, DaiS2018}. In addition, hBN has the ability to host a variety of single-photon emitters (SPEs) \cite{TranTT2016n, TranTT2016a, Jungwirth2016, Shotan2016, Kianinia2017, Grosso2017, Xia2019, Exarhos2019, LiuW2020, ChenY2020, Hayee2020} with extreme brightness \cite{Grosso2017, LiuW2020}, photoluminescence (PL) stability \cite{TranTT2016a, LiuW2020}, and large stark-shift tuning \cite{Shotan2016, Xia2019}. Most recently, some defects display spin-optical quantum properties and can be initialized, manipulated and optically read out at room temperature \cite{Gottscholl2020, Gottscholl2020a, LiuW2021aR, Chejanovsky2019, Mendelson2020, Stern2021}, significantly extending the functionality of hBN for quantum applications.

 \begin{figure*}[t]
  \centering
  \includegraphics[width=0.75\textwidth]{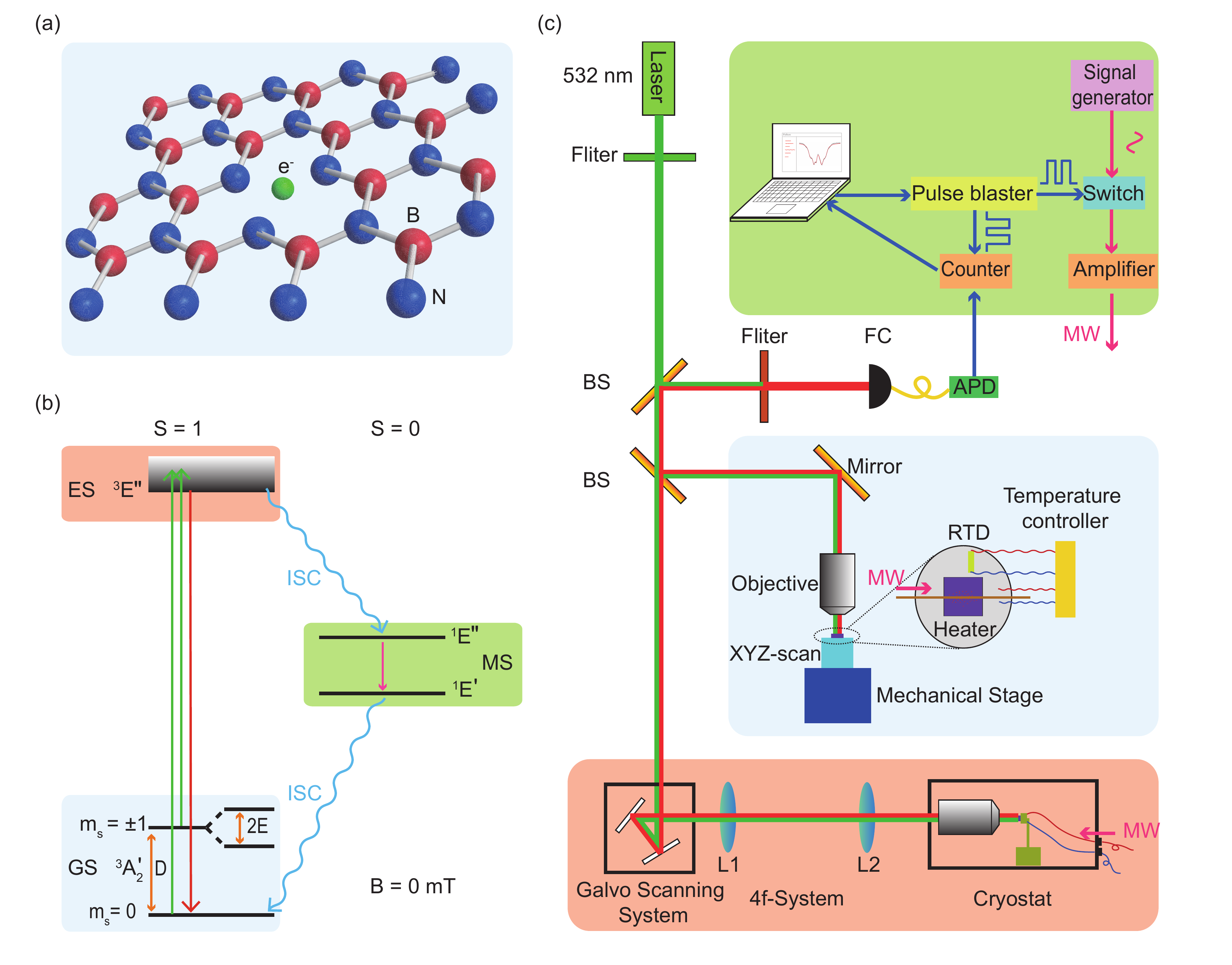}
  \caption{Structure of V$_\text{B}^-$ defect and experimental setup. (a) Geometrical structure of the V$_\text{B}^-$ defects in an hBN monolayer: a negatively charged boron vacancy surrounded by three equivalent nitrogen atoms (blue). (b) Simplified V$_\text{B}^-$ energy-level diagram and the transitions among the ground state ($^3A_2'$), the excited state ($^3E'$) and the metastable state ($^1E'$, $^1E''$) \cite{Ivady2020}. (c) Experimental setup of a homebuilt confocal microscopy system comprising a microwave system (green part), a heating system (blue part) and a cooling system (red part) for measuring the ODMR signal at different temperatures. BS: beam splitter; FC: fiber coupler; APD: avalanche photodiode; RTD: resistive temperature detector; MW: microwave.}\label{fig1}
\end{figure*}

Herein, we focus on one of the spin defects---the negatively charged boron vacancy (V$_\text{B}^-$) center, which comprises a missing boron atom and an extra electron surrounded by three equivalent nitrogen atoms in the h-BN lattice (Figure 1(a)). The highest point group symmetry of the defect is $\text{D}_{3\text{h}}$. Recent experimental and theoretical articles indicate that V$_\text{B}^-$ possesses a triplet ground state (S = 1) with a variety of low-energy triplet excited states (Figure 1(b)) \cite{Ivady2020, Sajid2020}. With $\lambda_\text{exc}$  = 532 nm laser excitation, we can observe strong emission spectra centered at $\sim$820 nm at temperatures from 5 to above 600 K.

The optically detected magnetic resonance (ODMR) method is a primary means for studying transitions between sublevels of the ground state. In this case, the V$_\text{B}^-$ defect will be polarized into the $\ket{m_s = 0}$ state by continuous laser irradiation (532 nm) due to spin-selective nonradiative intersystem crossings (ISCs) and pumped into the $\ket{m_s = \pm1}$ state and consequently exhibit a decrease in fluorescence intensity when the applied microwave (MW) energy is resonant with the split between ground state sublevels. The measured room-temperature ODMR spectrum of the V$_\text{B}^-$ ensemble embedded in hBN shows two distinct resonances, $\nu_1$ and $\nu_2$, located symmetrically around the frequency $\nu_0$, even without an external magnetic field. The parameters $D$ and $E$ that come from the spin-Hamiltonian (Eq. (1)) are often used to describe the zero-field splitting (ZFS) of the triplet ground state in solid color centers (e.g., NV$^-$ center in diamond \cite{Ivady2014}, divacancy center in SiC \cite{ZhouY2017, YanFF2020} and V$_\text{B}^-$ center in hBN \cite{Gottscholl2020, Gottscholl2020a}):
\begin{equation}
H = D(S_z^2-S(S+1)/3)+E(S_x^2-S_y^2)
\end{equation}
where $D$ and $E$ are the ZFS parameters, $S$ is the total electron spin ($S$=1 for V$_\text{B}^-$), the $z$-coordinate axis coincides with the $c$ axis of the hBN crystal, and $S_{x,y,z}$ are the spin-1 operators. The parameters can be derived from the ODMR spectrum as $D/h = \nu_0 = (\nu_1 +\nu_2)/2$ and $E/h = (\nu_2 - \nu_1)/2$, with $h$ being Planck's constant. Therefore, we can express the interaction energies in frequency units: $D = (\nu_1 +\nu_2)/2$ and $E = (\nu_2 - \nu_1)/2$.

While several papers have explored some of the properties \cite{Gottscholl2020, Gottscholl2020a, Abdi2018, Ivady2020, Sajid2020, Reimers2020} and methods of generating \cite{Toledo2018, Kianinia2020, GaoX2020} this defect, further experimental studies are needed for a better understanding. As with other color centers, the V$_\text{B}^-$ center shows great potential in quantum metrology, especially in temperature sensing due to the excellent thermal contact guaranteed by the two-dimensional nature of hBN. Moreover, we should consider the impact of environmental temperature fluctuations when V$_\text{B}^-$ defects are used for high-precision quantum control and ultrasensitive detection \cite{Gottscholl2021}. Therefore, studying the temperature-dependent features of this defect is of great significance.

In this work, we report the magnetic resonance spectra of the V$_\text{B}^-$ center in hBN nanopowders at different temperatures. Both the ZFS parameter $D$ and ODMR contrasts decrease as the temperature increases, while $E$ does not exhibit an obvious change. The $D$ shifts can be described by the Varshni equation very well and are approximately proportional to the reciprocal lattice volume. We speculate that thermal expansion is the main cause of energy-level shifts. We also roughly measure the distributions of $D$ and $E$ in our hBN nanopowdered samples, and our results indicate that there is significant difference among V$_\text{B}^-$ defects in different size hBN.

\begin{figure*}[t]
  \centering
  \includegraphics[width=0.75\textwidth]{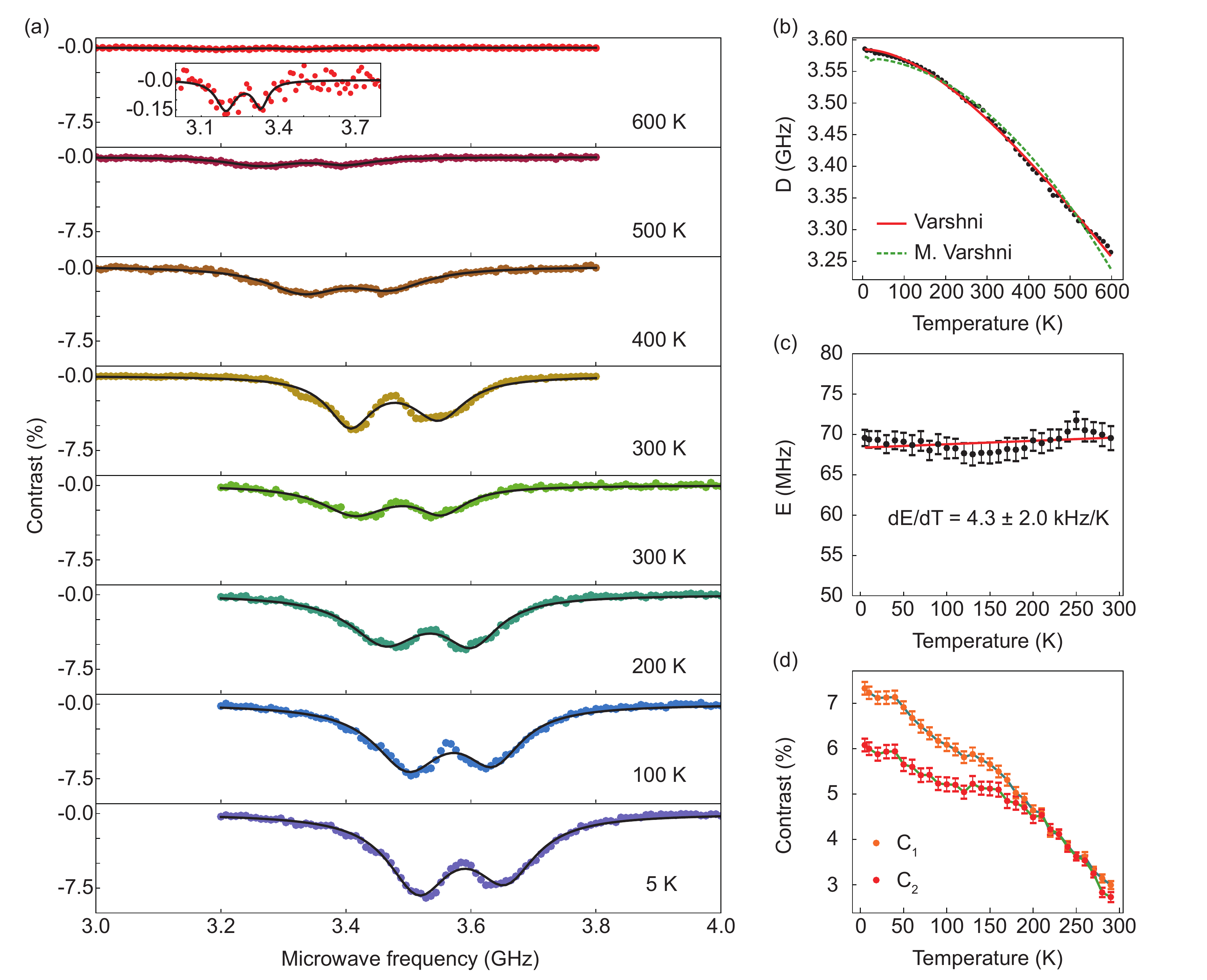}
  \caption{Temperature dependence of ODMR signals for V$_\text{B}^-$ defects. (a) ODMR spectra at different temperatures and the theoretical two-Lorentzian fits (black lines). The upper four pictures were measured in the heating system and the lower four were measured in the cooling system. The inset shows that the ODMR signal still exists, and the contrasts can reach 0.15\%, although it is very weak. (b) The ZFS parameter $D$ as a function of temperature from 5 to 600 K. The red line is the fit with the Varshni empirical equation, while the green dashed line is the fit with the modified Varshni equation. The error bar for each point is smaller than the corresponding symbol size. (c) The ZFS parameter $E$ plotted versus temperature with a linear fit (black line), which shows that it remains approximately constant. (d) The corresponding ODMR contrasts of the left branch (C$_1$) and right branch (C$_2$) as a function of temperature. In this article, we only consider variations in the absolute value of contrasts.}\label{fig2}
\end{figure*}

\textbf{Results and Discussion.} ODMR measurements were carried out by sweeping the frequency of the microwave field from 3.0 to 4.0 GHz, and the ODMR spectra at some representative temperatures with two-Lorentzian fits are shown in Figure 2(a). The upper four pictures were measured in the heating system and the lower four were measured in the cooling system. The excitation-laser intensity was 260 $\mu$W in the range 5-300 K (cooling system) and 170 $\mu$W in the range 300-600 K (heating system). At room temperature, the fluorescence signal dropped when the microwave field oscillated at $\nu_1$ $\sim$ 3.41 GHz and $\nu_2$ $\sim$ 3.55 GHz, which is consistent with the spin optical dynamics reported for V$_\text{B}^-$ defects in hBN \cite{Gottscholl2020}. Obviously, both the ZFS parameter $D$ and the contrasts decreased when the temperature increased, as shown in Figure 2(b) and (d). In this article, when considering variations in contrast, we only consider variations in the absolute values. The inset in Figure 2(a) shows that the ODMR signal persisted up to 600 K, and the contrasts reached approximately 0.15\%. Although the signal is very weak, this suggests that the V$_\text{B}^-$ defects still exist stably at a temperature above 600 K in air, and moreover, as reported in Ref. \cite{Toledo2018}, the defect can be maintained in an argon atmosphere until $\sim$750 K.

In Figure 2(b), the ZFS parameter $D$ has a nonlinear dependence on temperature from 5 to 600 K, and we fit it with the Varshni empirical equation \cite{Varshni1967}:
\begin{equation}
 D(T) = D_0 - \frac{\alpha T^2}{\beta+T}
\end{equation}
where $D_0$ is the value of $D$ at T = 0 K; $\alpha$ and $\beta$ are fitting parameter characteristics of the given material, which should be positive to ensure the monotonicity of $D$ as a function of temperature. $\alpha$ is an empirical constant, and $\beta$ is commonly supposed to be related and comparable to the Debye temperature $\Theta_\text{D}$. While this equation is regrettably very weak in its theoretical foundation and inconsistent with the Debye model in the low temperature limit, it is still one of the most widely quoted models due to the irreplaceable simplicity and monotonicity and has had great success in describing the temperature dependence of energy gaps in many semiconductor bulk and nanopowdered materials \cite{Onuma2016, Vainshtein1999, Odonnell1991, WangJ2012, Nepal2005, Passler1999, Vurgaftman2001}. In our experiment, the fitting based on the Varshni function showed outstanding agreement over the entire measured temperature range. As an example, the fit of the Varshni equation with parameters $D_0$ = 3584 $\pm$ 1 MHz, $\alpha$ = 1.06 $\pm$ 0.05 MHz/K, and $\beta$ = 559 $\pm$ 51 K in Figure 2(b) describes the $D$ shifts extremely well, with errors less than 12 MHz in the 5-600 K range. The value of $\beta$ is comparable to the Debye temperature, which is approximately 410 K \cite{Pease1952} in hBN. We also applied the third- \cite{Toyli2012, YanFF2018} and fifth- \cite{ZhouY2017} order polynomial functions to fit the experimental data. Both of them were slightly lacking in predictability at high temperature due to their nonmonotonicity (see Supporting Information Figure S2). Li et al. \cite{LiCC2017} proposed a modified Varshni equation to describe the energy-level shifts of NV$^-$ centers in diamond:
\begin{equation}
 D(T) = D_0 - \frac{AT^4}{(B+T)^2}
\end{equation}
where the two parameters $A$ and $B$ are restricted to positive values to ensure the monotonicity of the equation. However, the parameter $B$ obtained from the experimental data is negative, and the fitted modified Varshni equation (green dotted line in Figure 2(b)) is not monotonic in the low-temperature regime. This indicates that the temperature dependence of ZFS parameter $D$ of the V$_\text{B}^-$ center in hBN might be different from that of the NV$^-$ center in diamond, in which the electron-phonon interaction is the dominant mechanism for energy-level shifts \cite{Doherty2014, Ivady2014, LiCC2017}.

The ZFS parameter $E$ varies around 70 MHz between 5 and 300 K and is linearly fitted with a small slope of 4.3 $\pm$ 2.0 kHz/K (Figure 2(c)). This indicates that $E$ basically does not change with temperature, which shows the same behaviors of the NV$^-$ center in diamond \cite{Acosta2010} and the divacancy center in SiC \cite{ZhouY2017}. The dependence of the ODMR contrasts of the left branch (C$_1$) and right branch (C$_2$) on temperature is shown in Figure 2(d). The decreased contrasts might be caused by the thermally activated nonradiative processes in a manner similar to that described in Ref.\cite{Toyli2012}; following the operations described therein, and in order to get more relevant details, we measured the lifetime of the spin-resolved excited state with m$_s$ = 0 ($\tau_{m_s=0}$) in the temperature range 5-300 K. $\tau_{m_s=0}$($T$) in our case also shows a significant reduction with increasing temperature (see Supporting Information Figure S3), which might be the primary cause of the decrease in contrasts. The right branch contrast C$_2$ is smaller than the left branch contrast C$_1$ in the low-temperature regime. Further research is required to measure the lifetimes of excited states with $m_s = \pm 1$. We describe the measurement methods in the Supporting Information, Section S3, and it is difficult to measure the lifetimes $\tau_{m_s=\pm 1}$ with the current state of the art \cite{Gottscholl2020a, LiuW2021aR}.

 \begin{figure}[t]
  \centering
  \includegraphics[width=0.5\textwidth]{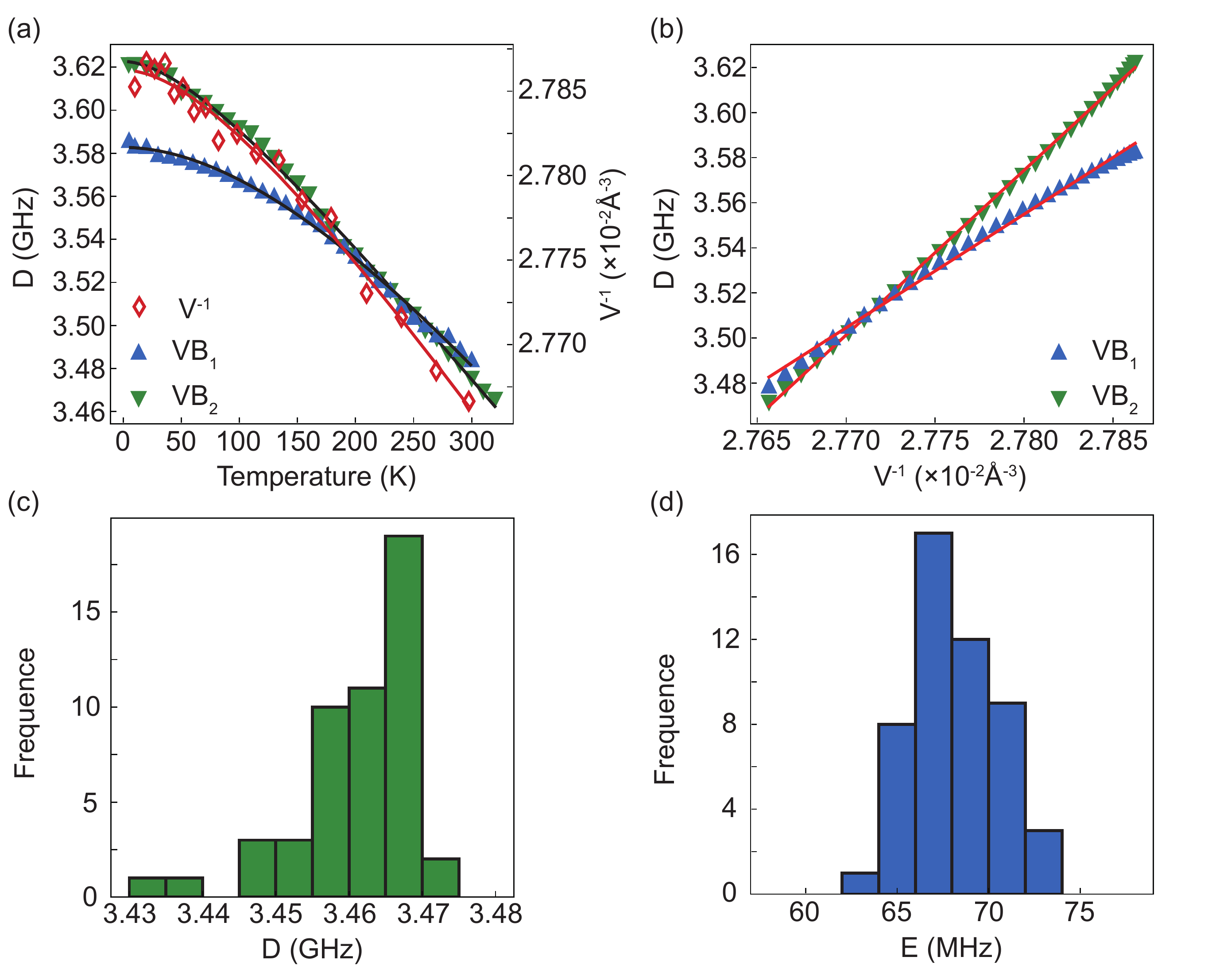}
  \caption{Difference in ODMR signals among hBN nanopowdered samples and relationship between $D$ and lattice volume. (a) Varshni empirical formula fitting for the reciprocal hBN lattice volume ($V^{-1}$) and the ZFS parameter $D$ of the hBN nanopowdered samples VB$_1$ and VB$_2$. The $V^{-1}$ curve represented by red diamonds was plotted by utilizing the hBN lattice-parameter data listed in Table 2 of Ref.\cite{Paszkowicz2002} by Paszkowicz et al. for the purpose of comparison. (b) The experimental data of $D$ of VB$_1$ and VB$_2$ were plotted as functions of $V^{-1}$. $V^{-1}$ is transformed from temperature using the fitting result in Figure 3(a). The red lines are the linear fits. (c) Histogram of ZFS parameter $D$ values from 50 V$_\text{B}^-$ centers under the same measurement conditions. The average $D$ value is 3460.8 $\pm$ 0.8 MHz. (d) Histogram of ZFS parameter $E$ values from V$_\text{B}^-$ centers from corresponding samples shown in (c). The average $E$ value is 68.2 $\pm$ 0.8 MHz. }\label{fig3}
\end{figure}

To reach a more general conclusion, we performed the same measurements on more hBN nanopowdered samples over the temperature range 5-300 K in the cooling system. The previous sample is labeled as VB$_1$, and the latter samples are VB$_{2-5}$ (VB$_2$ is also shown in the main text, and for VB$_{3-5}$, see Supporting Information, Section S5). The $D$ values for both VB$_1$ and VB$_2$ are shown in Figure 3(a) as a function of temperature. It is apparent that the overall decrease in $D$ with increasing temperature is consistent, and the curves were perfectly fitted by the Varshni empirical equation. There are still several differences, e.g., $D$ of VB$_2$ is larger below 200 K, $E$ of VB$_2$ is approximately 80 MHz instead of 70 MHz as with VB$_1$ (see Supporting Information Figure S7(a)). Furthermore, we measured the ODMR signals of fifty other hBN samples at room temperature under the same experimental conditions to perform a statistical comparison of different nanopowders (Figure 3(c,d)). The $D$ values with an average of 3460.8 $\pm$ 0.8 MHz were unevenly distributed between 3.43 and 3.47 GHz, and the $E$ values with an average of 68.2 $\pm$ 0.8 MHz were approximately normally distributed between 62 and 74 MHz. It is worth noting that the $D$ of nanopowders is approximately consistent with the $D$ of bulk hBN measured in previous experiments, but $E$ is quite different from the previously observed 50 MHz even when the fitting error is taken into account \cite{Gottscholl2020, Kianinia2020, GaoX2020}. The $E$ values we measured were close to the 80 MHz value calculated in Ref.\cite{Sajid2020}. We speculate that the local strain in the hBN nanopowders and the distance between defects and flake edges are the underlying causes of the difference in different samples \cite{Hayee2020, Sajid2020} (see Supporting Information Sections S1 and S4 for more details and discussion). We also think that the impurities and surface morphology are other possible causes of the difference, and more theoretical and experimental work is needed.

It is interesting that the significant difference in $D$ shifts of the V$_\text{B}^-$ center in different hBN nanopowders is inconsistent with that of the NV$^-$ center in diamond, as discussed in Refs.\cite{Acosta2010, Doherty2014}. We conjecture that the dominant contribution to energy-level shifts is thermal expansion rather than electron-phonon interactions. Analogous to the calculations in diamond \cite{Acosta2010, ChenXD2011, Doherty2014, Ivady2014}, $D$ is mainly determined by electron spin-spin interactions, which can be simply written as $D$ $\propto$ $\langle1/r^3\rangle$, where $r$ is the displacement between the two dipoles. Furthermore, we assume that $D$ $\propto$ $\langle V^{-1}\rangle$, where $V$ is the lattice volume. Paszkowicz et al. measured the lattice parameters of hBN in the temperature range 10-297.5 K, and the original data are listed in Table 2 in Ref.\cite{Paszkowicz2002}. Utilizing these data, we plotted the $V^{-1}$-temperature curve in the same figure for comparison (see Figure 3(a)) and fit it with the Varshni equation. Then we present the $D-V^{-1}$ relationship in Figure 3(b), where the horizontal coordinate $V^{-1}$ was calculated from the set temperature using this fitting result. The linear fits with the slope of 500.02 $\pm$ 8.34 GHz$\cdot$\AA$^3$ and 727.90 $\pm$ 6.94 GHz$\cdot$\AA$^3$, respectively, indicated that the $D$ shifts are indeed proportional to the reciprocal lattice volume, which supports our previous hypothesis. The two different slopes may be used to characterize the different hBN nanopowders. Notably, the anisotropic thermal expansion of hBN \cite{Paszkowicz2002} should be considered, and further experimental, and computational studies are needed in the future. Additionally, more properties of the parameter D for our sample (including that depends on the excitation-laser intensity) are found in the Supporting Information.

To summarize, we measured the temperature dependence of the ZFS parameters of V$_\text{B}^-$ defects in hBN nanopowders and concluded that thermal expansion is the dominant cause.

\textbf{Methods.} \emph{Sample Preparation.} HBN ultrafine powder with 99.0\% purity and $\sim$70 nm particle size were obtained commercially from Graphene Supermarket and neutron irradiated with an integrated thermal flux of 7$\times$10$^{17}$ $n$ cm$^{-2}$ in a nuclear reactor. A small amount of hBN nanopowders was mixed in hydrogen peroxide (H$_2$O$_2$) solvent, sonicated into a suspension, and finally drop-cast onto the silicon (Si) substrate with a thermally grown SiO$_2$ top layer ($\sim$300 nm).

\emph{Experimental Setup.} A homebuilt confocal microscopy system comprising a microwave system, a heating system and a cooling system was used for the ODMR measurements at different temperatures (Figure 1(c)). A 532-nm continuous-wave (cw) laser filtered by a 532-nm bandpass filter (LL01-532-25, Semrock) was focused on the samples through a N.A. = 0.9 objective (Olympus MPLFLN100xBDP) for spin initialization and excitation. The fluorescence was collected by the same objective, and the portion filtered by the 750-nm long-pass filter (FELH0750, Thorlabs) was guided through a 9-$\mu$m-core-diameter fiber to an avalanche photodiode (APD) for signal readout. In the microwave system, the microwave irradiation generated by a synthesized signal generator (SSG-6000RC, Mini-Circuits) went through a microwave switch and an amplifier, and then a 20-$\mu$m diameter copper wire suspended in close proximity ($\sim\mu$m away) to the samples served as an antenna to deliver the microwave field. A 500-MHz pulse blaster card was used to generate the electrical pulse sequences for manipulating the microwave and reading out the signal. In the heating system, on-chip heating and thermometry elements that combined a metal ceramic heater and a resistive temperature detector (RTD) were attached to a temperature controller to achieve stability within $\pm$ 100 mK at temperatures up to 600 K in air. In the cooling system, the cryogenic chamber (PSE-237 Cryostation s200-CO, Montana Instruments) had its own temperature controller (Model 335, Lake Share), which stabilized the temperature to $\pm$ 1 mK.

\section*{Acknowledgments}

We are grateful to the helpful discussions of Igor Aharonovich, Klaus Kramborck and Jos\'{e} Roberto de Toledo on the sample fabrication. This work is supported by the National Key Research and Development Program of China (No. 2017YFA0304100), the National Natural Science Foundation of China (Grants Nos. 11822408, 11674304, 11774335, 11821404, and 11904356), the Key Research Program of Frontier Sciences of the Chinese Academy of Sciences (Grant No. QYZDY-SSW-SLH003), the Fok Ying-Tong Education Foundation (No. 171007), the Youth Innovation Promotion Association of Chinese Academy of Sciences (Grants No. 2017492), Science Foundation of the CAS (No. ZDRW-XH-2019-1), Anhui Initiative in Quantum Information Technologies (AHY020100, AHY060300), the Fundamental Research Funds for the Central Universities (Nos. WK2470000026, WK2030000008 and WK2470000028).

\section*{Supporting Information}
This material is available free of charge via the internet at http://pubs.acs.org. 

   Full experimental data, fitting results, and excitation-laser intensity dependence.

\end{document}